\documentclass[twocolumn]{aastex63}

\usepackage{amsmath}	
\usepackage{enumitem}
\usepackage[space]{grffile}  
\usepackage{bm} 
\usepackage{makecell} 

\def\Kepler{\textit{Kepler}} 
\def\Gaia{\textit{Gaia}} 

\def\SysSim{\textit{SysSim}} 

\defcitealias{2020AJ....160..276H}{H20}

\received{}
\revised{}
\accepted{}
\shorttitle{Debiasing the Minimum-Mass Extrasolar Nebula}
\shortauthors{He \& Ford}
\graphicspath{{./}{Figures/}}

\begin{document}

\title{Debiasing the Minimum-Mass Extrasolar Nebula: On the Diversity of Solid Disk Profiles}

\correspondingauthor{Matthias Yang He}
\email{mhe@nd.edu}

\author[0000-0002-5223-7945]{Matthias Y. He}
\affiliation{Department of Physics \& Astronomy, 225 Nieuwland Science Hall, The University of Notre Dame, Notre Dame, IN 46556, USA}
\affiliation{Department of Astronomy \& Astrophysics, 525 Davey Laboratory, The Pennsylvania State University, University Park, PA 16802, USA}
\affiliation{Center for Exoplanets \& Habitable Worlds, 525 Davey Laboratory, The Pennsylvania State University, University Park, PA 16802, USA}
\affiliation{Center for Astrostatistics, 525 Davey Laboratory, The Pennsylvania State University, University Park, PA 16802, USA}
\affiliation{Institute for Computational \& Data Sciences, 525 Davey Laboratory, The Pennsylvania State University, University Park, PA 16802, USA}

\author[0000-0001-6545-639X]{Eric B. Ford}
\affiliation{Department of Astronomy \& Astrophysics, 525 Davey Laboratory, The Pennsylvania State University, University Park, PA 16802, USA}
\affiliation{Center for Exoplanets \& Habitable Worlds, 525 Davey Laboratory, The Pennsylvania State University, University Park, PA 16802, USA}
\affiliation{Center for Astrostatistics, 525 Davey Laboratory, The Pennsylvania State University, University Park, PA 16802, USA}
\affiliation{Institute for Computational \& Data Sciences, 525 Davey Laboratory, The Pennsylvania State University, University Park, PA 16802, USA}
\affiliation{Institute for Advanced Study, 1 Einstein Drive, Princeton, NJ 08540, USA}
\affiliation{Center for Computational Astrophysics, Flatiron Institute, New York, NY 10010, USA}



\begin{abstract}

A foundational idea in the theory of \textit{in situ} planet formation is the ``minimum mass extrasolar nebula'' (MMEN), a surface density profile ($\Sigma$) of disk solids that is necessary to form the planets in their present locations.
While most previous studies have fit a single power-law to all exoplanets in an observed ensemble, it is unclear whether most exoplanetary systems form from a universal disk template.
We use an advanced statistical model for the underlying architectures of multi-planet systems to reconstruct the MMEN.
The simulated physical and \Kepler{}-observed catalogs allow us to directly assess the role of detection biases, and in particular the effect of non-transiting or otherwise undetected planets, in altering the inferred MMEN.
We find that fitting a power-law of the form $\Sigma = \Sigma_0^* (a/a_0)^\beta$ to each multi-planet system results in a broad distribution of disk profiles; $\Sigma_0^* = 336_{-291}^{+727}$ g/cm$^2$ and $\beta = -1.98_{-1.52}^{+1.55}$ encompass the 16th-84th percentiles of the marginal distributions in an underlying population, where $\Sigma_0^*$ is the normalization at $a_0 = 0.3$ AU.
Around half of the inner planet-forming disks have minimum solid masses of $\gtrsim 40 M_\oplus$ within 1 AU.
While transit observations do not tend to bias the median $\beta$, they can lead to both significantly over- and under-estimated $\Sigma_0^*$ and thus broaden the inferred distribution of disk masses.
Nevertheless, detection biases cannot account for the full variance in the observed disk profiles; there is no universal MMEN if all planets formed \textit{in situ}.
The great diversity of solid disk profiles suggests that a substantial fraction ($\gtrsim 23\%$) of planetary systems experienced a history of migration.

\end{abstract}



\section{Introduction} \label{sec:Intro}


Well before the discovery of extrasolar planets, the idea of a ``minimum mass'' solar nebula (MMSN) was introduced to posit the \textit{in situ} formation of the planets in our solar system \citep{1977Ap&SS..51..153W, 1985prpl.conf.1100H}. The MMSN framework is intuitively simple: in this view, the initial proto-planetary disk must have enough material to form the final planets in their present locations, and in particular must contain enough solids for the planets to locally accrete up to their core masses within their feeding zones. This implies that one can work backwards to infer the requisite (solid) surface density, $\Sigma$, local to each planet by spreading its mass in an annulus centered around its semi-major axis ($a$). By assuming a smooth disk profile, a power-law for $\Sigma$ as a function of $a$ is typically fit to then infer the radial distribution of disk solids.

Since the discovery of thousands of exoplanets, largely due to the transformational success of NASA's \Kepler{} mission, numerous studies have applied the MMSN template to these extrasolar worlds in order to form an analogous minimum mass \textit{extrasolar} nebula (MMEN; \citealt{2013MNRAS.431.3444C, 2014MNRAS.440L..11R, 2014ApJ...795L..15S, 2020AJ....159..247D}).
Yet, it continues to be debated whether most exoplanetary systems conform to a universal disk profile.
Most previous studies have fit a single disk profile, in the form of a single power-law for $\Sigma(a)$, to all the exoplanet candidates observed by \Kepler{} simultaneously \citep{2013MNRAS.431.3444C, 2014ApJ...795L..15S, 2020AJ....159..247D}.
This construction has multiple shortcomings: (1) it washes out any potential system-level correlations, (2) the resulting ``MMEN'' does not represent the properties of any single true/physical disk, and (3) it does not account for planet multiplicity, and specifically the effect of missing (undetected) planets in each system.
One notable exception\footnote{A pioneering study by \citet{2004ApJ...612.1147K} also fit power-laws to individual planetary systems, but with exoplanets detected by the radial velocity method and were limited to a much smaller sample of systems.} is \citet{2014MNRAS.440L..11R} (hereafter RC14), who fit a power-law to each individual system with three or more observed planets. They showed that this produces a broad diversity of minimum-mass disks with profiles ranging from $\Sigma \propto a^{-3.2}$ to $a^{0.5}$, thus retaining the variances across individual systems and illustrating the inconsistency of a universal disk profile.

Previous studies have also relied on simplistic treatments for detection biases and the exoplanet mass-radius relationships to construct the MMEN from the \Kepler{} planet catalog, typically by applying a correction factor for the transit geometric and detection probability (i.e., a form of inverse detection efficiency) of each planet in an attempt to ``debias'' the observed sample \citep{2013MNRAS.431.3444C, 2020AJ....159..247D}.
While this approach effectively weights the planets such that longer period and smaller sized planets are compensated for their reduced detectability by transits, it is a clear oversimplification of the \Kepler{} detection pipeline \citep{2020AJ....160..159C} and does not correct for missing planets in systems with known planet(s).
The recent development of detailed forward models for the \Kepler{} mission has enabled unprecedented inferences on the intrinsic population of inner planetary systems from the observed population (e.g., \citealt{2019MNRAS.490.4575H}), leading to advanced statistical models such as the ``maximum AMD model'' that captures the underlying architectures and correlations in multi-planet systems \citep{2020AJ....160..276H}.
The \SysSim{} simulated catalogs, comprised of \textit{physical} and \textit{observed} catalog pairs, provide a way to directly address the impact of non-transiting or otherwise unseen planets in \Kepler{}-like systems.

In this article, we use simulated catalogs (physical and observed) to assess how detection biases can affect our interpretation of the MMEN and make comparisons to the \Kepler{} observed catalog.
In \S\ref{sec:Methods}, we describe how we build the MMEN, beginning with a summary of the \SysSim{} population model developed previously to reproduce the \Kepler{} planet catalog (\S\ref{sec:AMD_model}), and a review of how planets are converted to solid surface densities using various prescriptions for their feeding zones found in the literature (\S\ref{sec:surface_densities}).
We discuss the standard method in which a power-law is fit to all the exoplanets in a given catalog, and show differences in fitting to the simulated observed and physical catalogs (\S\ref{sec:fit_all}).
We then adopt a procedure modified from that of RC14 in which a power-law disk profile is fit to each multi-planet system to directly construct the distribution of MMEN from the physical catalogs (\S\ref{sec:fit_systems}).
In \S\ref{sec:missing_planets}, this procedure is repeated for observed multi-transiting systems (simulated and \Kepler{}) to assess the effect of missing planets in altering the inferred MMEN distribution.
We discuss the implications for the distribution of minimum disk masses in \S\ref{sec:disk_masses}.
Finally, we summarize and discuss our key results in \S\ref{sec:discussion}.

\section{Constructing the Minimum Mass Extrasolar Nebula} \label{sec:Methods}


\subsection{Population model} \label{sec:AMD_model}

\begin{figure*}
\centering
\includegraphics[scale=0.46,trim={0.5cm 1cm 0.5cm 1cm},clip]{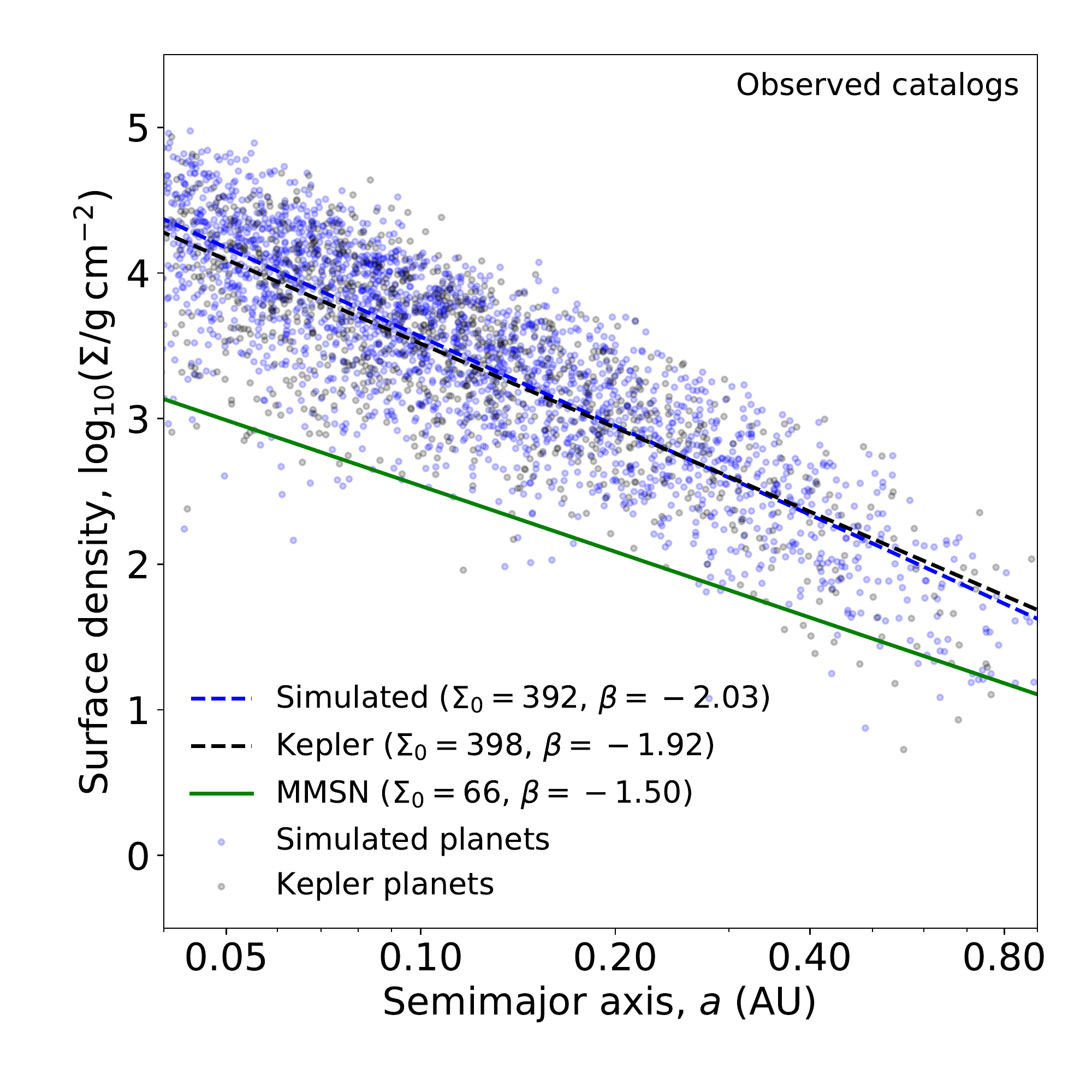}
\includegraphics[scale=0.46,trim={0.5cm 1cm 0.5cm 1cm},clip]{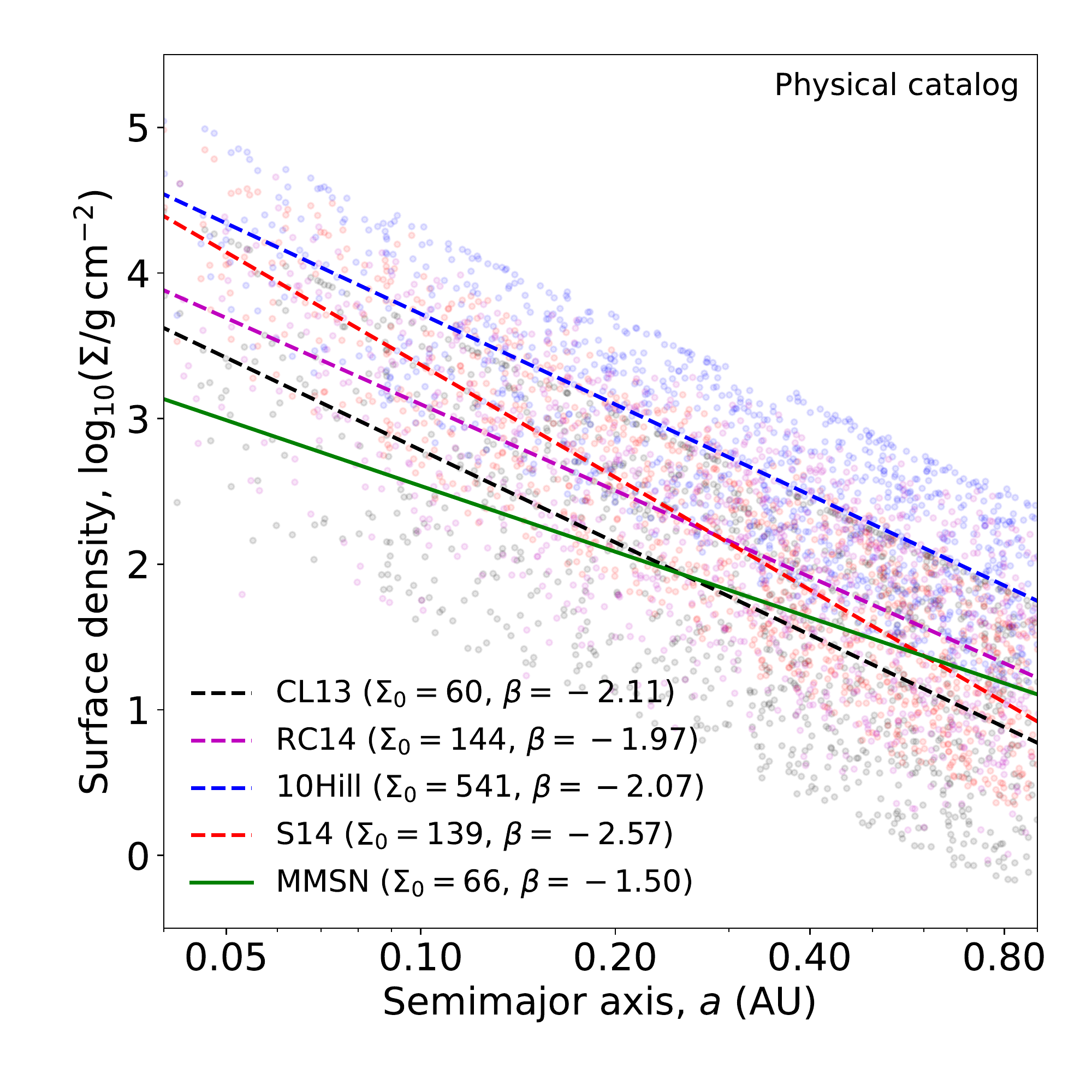} 
\caption{\textbf{Left:} solid surface densities vs. semi-major axes for \Kepler{} planet candidates (black) and simulated-observed planets (blue) in multi-transiting systems. The RC14 prescription is used here for all planets. The dashed lines show the corresponding power-law fits over the observed planets. For comparison, the solid green line shows an estimate of the minimum mass solar nebula (MMSN; equation 2 of \citealt{2010AREPS..38..493C}). For each power-law fit, $\beta$ represents the slope while $\Sigma_0$ (given in g/cm$^2$) represents the normalization at $a_0 = 0.3$ AU (equation \ref{eq_mmen}). The power-law fit parameters using the other $\Delta{a}$ prescriptions are provided in Table \ref{tab:params}.
\textbf{Right:} same as the left panel, but with simulated planets from a physical catalog, for various prescriptions of $\Delta{a}$ (as labeled). Each set of colored points denotes the same sample of simulated planets, with surface densities computed using a given $\Delta{a}$ prescription. Likewise, each colored dashed line denotes a power-law fit over all planets (observed or not) in the catalog. We note that the RC14 prescription (purple) can only be computed for systems with at least two planets.}
\label{fig:mmen_deltaa}
\end{figure*}

The ``maximum AMD model'' (\citealt{2020AJ....160..276H}; hereafter, the \citetalias{2020AJ....160..276H} model) was developed to describe as many features of the \Kepler{} planet catalog as possible, with a focus on the correlated properties of planets in multi-transiting systems, using a combination of statistical distributions and conditions for dynamical stability \citep{2017A&A...605A..72L, 2017A&A...607A..35P}.
It provides a detailed parameterization of the underlying distribution of planetary systems between orbital periods of $P = 3 - 300$ days and planet sizes of $R_p = 0.5 - 10 R_\oplus$ around a purified sample of FGK dwarf stars (see \citealt{2019AJ....158..109H} and \citetalias{2020AJ....160..276H} for a description of the stellar selection criteria).
The \Kepler{} planetary systems in this sample were derived from the Kepler Objects of Interest DR25 data set obtained from the NASA Exoplanet Archive \citep{koidr25}\footnote{Accessed on 2020-10-19 at 22:34.} and include 2169 planet candidates of which 964 are in 388 multi-transiting systems.
For this planet sample, the model reproduces an astounding number of properties at the population level (see \citetalias{2020AJ....160..276H}; \citealt{2021AJ....161...16H}; \citealt{2021AJ....162..166M}), including but not limited to the following which are most relevant to this study: (1) the overall number of planets per star and the observed multiplicity distribution, (2) the period and period ratio distributions, and (3) the intra-system size similarity patterns, of which the latter two are commonly referred to as the ``peas-in-a-pod'' patterns (see \citealt{2022arXiv220310076W} for a review).

\emph{Mass-radius relation:} a planetary mass-radius (M-R) relation is necessary to estimate planet masses from their radii. The \citetalias{2020AJ....160..276H} model uses a probabilistic M-R relation that consists of a lognormal distribution centered around the ``Earth-like rocky'' model from \citet{2019PNAS..116.9723Z} for small planets and a non-parametric model defined by a series of Bernstein polynomials fit to a sample of 127 \Kepler{} planets with RV or TTV masses from \citet{2018ApJ...869....5N} for large planets, with the transition radius from small to large chosen to be $1.472 R_\oplus$ such that both the mean prediction and scatter are continuous across the entire range of radii.
This M-R relation is thus motivated by a combination of both physical (below the transition radius) and empirical (above the transition radius) models and is more detailed than previously adopted relations for computing the MMEN from the \Kepler{} planets.

\SysSim{} enables the generation of physical and observed catalog pairs from the \citetalias{2020AJ....160..276H} model (the code is available at \citealt{eric_ford_2022_5915004, matthias_yang_he_2022_5963884}).
A physical catalog represents (one realization of) the true, underlying distribution of planetary systems.
Each system in this catalog consists of a known \Kepler{} target star (with stellar parameters from the \Gaia{}-\Kepler{} Stellar Properties Catalog; \citealt{2020AJ....159..280B}) and a set of planets with physical radii, masses, orbital periods, and orbital elements drawn directly from the \citetalias{2020AJ....160..276H} model.
An observed catalog then represents one realization of the detected transiting planets from the physical catalog under a \Kepler{}-like (primary) mission, simulated using a detailed model for the combined detection and vetting efficiency of the \Kepler{} pipeline that accounts for the window and 1-$\sigma$ depth functions of each individual target star (\citealt{2017ksci.rept...14B, 2020AJ....160..159C}; see \citealt{2019AJ....158..109H} for details).

\begin{deluxetable*}{lcccccccc}
\centering
\tablecaption{Power-law fits (equation \ref{eq_mmen}, with $a_0 = 0.3$ AU) for various prescriptions of $\Delta{a}$ and ensembles of planets.}
\tablehead{
 \colhead{} & \multicolumn2c{CL13} & \multicolumn2c{RC14} & \multicolumn2c{10Hill} & \multicolumn2c{S14} \\
 \colhead{Fit all planets} & \colhead{$\Sigma_0$ (g/cm$^2$)} & \colhead{$\beta$} & \colhead{$\Sigma_0$ (g/cm$^2$)} & \colhead{$\beta$} & \colhead{$\Sigma_0$ (g/cm$^2$)} & \colhead{$\beta$} & \colhead{$\Sigma_0$ (g/cm$^2$)} & \colhead{$\beta$}
}
\startdata
 Physical catalogs & $65_{-8}^{+9}$ & $-2.08_{-0.03}^{+0.02}$ & $146_{-15}^{+16}$ & $-1.97_{-0.03}^{+0.05}$ & $568_{-45}^{+47}$ & $-2.05_{-0.02}^{+0.01}$ & $147_{-12}^{+13}$ & $-2.55_{-0.02}^{+0.01}$ \\[3pt]
 Observed catalogs & $246_{-6}^{+7}$ & $-1.84_{-0.02}^{+0.04}$ & $369_{-27}^{+42}$ & $-2.03_{-0.05}^{+0.05}$ & $1396_{-33}^{+22}$ & $-1.89_{-0.02}^{+0.02}$ & $414_{-10}^{+12}$ & $-2.35_{-0.02}^{+0.02}$ \\[3pt]
 Kepler catalog & $264_{-5}^{+5}$ & $-1.67_{-0.02}^{+0.02}$ & $413_{-13}^{+13}$ & $-1.90_{-0.03}^{+0.03}$ & $1466_{-17}^{+20}$ & $-1.77_{-0.01}^{+0.01}$ & $445_{-4}^{+4}$ & $-2.18_{-0.01}^{+0.01}$ \\[3pt]
 \hline
 \rule{0pt}{4ex}\makecell{Fit each system} & \makecell{$\Sigma_0^*$ (g/cm$^2$)} & \makecell{$\beta$} & \makecell{$\Sigma_0^*$ (g/cm$^2$)} & \makecell{$\beta$} & \makecell{$\Sigma_0^*$ (g/cm$^2$)} & \makecell{$\beta$} & \makecell{$\Sigma_0^*$ (g/cm$^2$)} & \makecell{$\beta$} \\[3pt]
 \hline
 Physical catalog & $160_{-140}^{+313}$ & $-2.02_{-1.51}^{+1.50}$ & $336_{-291}^{+727}$ & $-1.98_{-1.52}^{+1.55}$ & $1038_{-771}^{+1125}$ & $-2.02_{-1.00}^{+1.00}$ & $263_{-196}^{+330}$ & $-2.53_{-0.99}^{+1.00}$ \\[3pt]
 Observed catalog & $349_{-284}^{+597}$ & $-1.99_{-1.28}^{+1.36}$ & $519_{-420}^{+2100}$ & $-1.97_{-1.27}^{+1.54}$ &  $1758_{-1214}^{+1766}$ & $-2.00_{-0.82}^{+0.96}$ & $496_{-328}^{+539}$ & $-2.39_{-0.79}^{+0.85}$ \\[3pt]
 Kepler catalog & $388_{-308}^{+914}$ & $-1.84_{-1.25}^{+1.52}$ & $515_{-428}^{+2255}$ & $-1.80_{-1.28}^{+1.59}$ & $1776_{-1048}^{+2790}$ & $-1.82_{-0.72}^{+1.01}$ & $519_{-332}^{+723}$ & $-2.25_{-0.80}^{+0.91}$ \\[3pt]
\enddata
\tablecomments{The uncertainties for ``Fit all planets'' denote the 16th-84th percentiles computed over many iterations of simulated catalogs (each with the same number of target stars as the \Kepler{} catalog), while the uncertainties for ``Fit each system'' represent the 16th-84th percentiles of the distributions over all the systems in a single catalog. Since there is only one \Kepler{} catalog, the uncertainties for ``Fit all planets'' in the \Kepler{} catalog were computed via re-samplings of the mass-radius relation for all of the planets.}
\tablenotetext{*}{The fitted values of $\Sigma_0$ have been scaled up by a factor $\alpha$ for each system, as described in \S\ref{sec:fit_systems}.}
\label{tab:params}
\end{deluxetable*}

\subsection{Computing solid surface densities from planets} \label{sec:surface_densities}

The minimum solid surface density ($\Sigma$) required to form a planet is given by spreading the mass of the planet in solids ($M_p$) over an annulus of width $\Delta{a}$ (which conceptually represents the feeding zone of the planet) centered at a separation of $a$:
\begin{equation}
 \Sigma = \frac{M_p}{2\pi{a}\Delta{a}}. \label{eq_ssd}
\end{equation}
Since the solid mass of a planet is limited to its core mass and most planets above $\sim 1.6 R_\oplus$ have gaseous envelopes \citep{2015ApJ...801...41R}, we also limit $M_p$ to $10 M_\oplus$. The choice of setting $10 M_\oplus$ as the maximum core mass is motivated by numerous studies on the critical core mass for runaway gas accretion \citep{1982P&SS...30..755S, 1996Icar..124...62P, 2006ApJ...648..666R, 2014ApJ...797...95L, 2015ApJ...800...82P}, which find $M_{\rm core} \simeq 5-20 M_\oplus$. Yet, it is possible for some planets to have solid masses greater than 10 Earth masses (potentially including Jupiter and Saturn; \citealt{1999P&SS...47.1183G, 2017GeoRL..44.4649W, 2019Natur.572..355L}); thus, we also repeat our analyses without setting any maximum core mass and find little change in our results\footnote{For example, while the upper tail of our distribution in solid surface density normalizations (later defined in \S\ref{sec:fit_systems}) extends to modestly higher values, the median value increases by only $\sim 2\%$. Thus, our results are insensitive to our assumption of the exact core mass limit.}, due to the infrequency of large/giant planets in the \citetalias{2020AJ....160..276H} model.
For each planet, we then compute $\Sigma$ using a number of previously adopted prescriptions for the feeding zone width $\Delta{a}$:

\begin{enumerate}
 \item The simplest prescription is to adopt a width equal to the semi-major axis (\citealt{2013MNRAS.431.3444C}; hereafter CL13),
  \begin{equation}
   \Delta{a} = a. \label{eq_deltaa_CL2013}
  \end{equation}
 While convenient, this prescription tends to result in overlapping regions between planets in the same system and likely overestimates $\Delta{a}$, thus underestimating $\Sigma$.
  
 \item RC14 recommend using the geometric means of the semi-major axes for neighboring planets as the dividing boundaries for their feeding zones,
  \begin{align}
   a_{{\rm sep},i} &= \sqrt{a_i a_{i+1}}, \quad i = 1,\dotsc,N-1 \\
   \Delta{a_i} &= a_{{\rm sep},i} - a_{{\rm sep},i-1}, \label{eq_deltaa_RC2014}
  \end{align}
  where $a_{{\rm sep},i}$ is the boundary between the $i$ and $i+1$ planets, $\Delta{a_i}$ is the width for the $i^{\rm th}$ planet, and $N$ is the number of (physical or observed) planets in the system. For the inner (outer) edge of the innermost (outermost) planet, we define it by enforcing the same ratio in $a$ both interior and exterior to the planet. We note that this prescription is only applicable to multi-planet systems and may overestimate (or underestimate) $\Delta{a}$ when not all planets are detected.
  
 \item Alternatively, one can justify using a multiple of the planet's Hill radius:
  \begin{equation}
   \Delta{a} = k R_{\rm Hill} = k{a}\Big(\frac{M_p}{3M_\star}\Big)^{1/3}, \label{eq_deltaa_kHill}
  \end{equation}
  where $k$ is a constant factor and $M_\star$ is the mass of the host star. \citet{2020AJ....159..247D} chose $k = 10$ as motivated by the spacings of \Kepler{} observed multi-planet systems \citep{2018AJ....155...48W}, which we denote hereafter as 10Hill. While this is appropriate for low-mass planets on low-eccentricity orbits, it is expected to underestimate the width of the feeding zone for systems with significant eccentricities.
  
 \item Finally, \citet{2014ApJ...795L..15S} (hereafter S14) suggests
  \begin{equation}
   \Delta{a} = 2^{3/2} a \sqrt{\frac{a M_p}{R_p M_\star}}, \label{eq_deltaa_S2014}
  \end{equation}
   motivated by considerations for the role of giant impacts in dictating a planet's effective feeding zone width.
\end{enumerate}

In Figure \ref{fig:mmen_deltaa} (left panel), we plot the solid surface density versus semi-major axis for each planet in the \Kepler{} catalog (black points) as well as in a simulated observed catalog (blue points). To facilitate the rest of the paper, the RC14 prescription is used for the purposes of this panel.
Similarly, we plot the solid surface densities for a sample of $10^3$ simulated planets drawn from a simulated physical catalog (right panel). Each planet is repeated as four points (once for each of the above prescriptions). While there is significant scatter in $\Sigma$ due to both (1) the range of planet masses (up to two orders of magnitude even after capping the core masses; $M_p \sim 0.1-10 M_\oplus$) and (2) the varying prescriptions for $\Delta{a}$, there is a clear linear trend in $\log{\Sigma}$ vs. $\log{a}$ indicative of a power-law relation between $\Sigma$ and $a$ that is qualitatively consistent with previous studies. In the next subsection, we describe our procedure for fitting power-laws to both the total ensemble of planets as well as individual multi-planet systems.

\subsection{The canonical power-law model} \label{sec:fit_all}

A power-law model for the MMEN is typically fitted to the solid surface densities computed from the planets as a function of the semi-major axis, of the form:
\begin{equation}
 \Sigma(a) = \Sigma_0 \bigg(\frac{a}{a_0}\bigg)^\beta, \label{eq_mmen}
\end{equation}
where $\Sigma_0 \equiv \Sigma(a_0)$ is the normalization at separation $a_0$ and $\beta$ is the slope. While $a_0$ is typically assumed to be 1 AU, we choose $a_0 = 0.3$ AU so that it is closer to the median $a$ of the simulated and \Kepler{} planets, and thus reduces the covariance of $\Sigma_0$ and $\beta$.
Most previous studies have fitted equation \ref{eq_mmen} to an ensemble of planets (from transit or RV surveys) to construct a single, ``universal'' MMEN. To facilitate direct comparisons with these prior studies, we also fit a power-law to all the planets in each of the catalogs (\Kepler{} observed, simulated observed, and simulated physical), as denoted by the various dashed lines in Figure \ref{fig:mmen_deltaa}. The complete results for all prescriptions of $\Delta{a}$ are presented in Table \ref{tab:params} under ``Fit all planets''.

First, we compare the results of the simulated observed catalogs to the \Kepler{}-observed catalog. While there is a qualitatively good agreement between the two, the fits to the simulated planets give slightly lower normalizations at 0.3 AU ($\sim 5-10\%$ smaller values of $\Sigma_0$) for all prescriptions. Similarly, the values of $\beta$ are slightly steeper in the simulated observed catalogs than in the actual \Kepler{} observations.
These differences are likely due to having fewer simulated systems with positive size ordering (``monotonicity''; see Figure 12 of \citetalias{2020AJ....160..276H}) compared to the \Kepler{} data, i.e. there are slightly more larger planets at short periods relative to the \Kepler{} systems.
However, while these differences are statistically significant over many simulated catalogs, they are smaller than the differences arising from the various prescriptions.

Next, we compare how the fits to the simulated planets change between the physical and observed catalogs. This comparison quantifies how the detection biases of the transit survey alter the inferred (mean) MMEN. We find that the values of $\Sigma_0$ are a factor of $\sim 2.5-4$ lower for the physical catalogs compared to the observed catalogs; this is readily explained by the fact that transit detections are biased towards larger planets (which would tend to be more massive) at all separations. Interestingly, $\beta$ is only modestly affected, appearing to be slightly dampened by the detection biases; surprisingly, the opposite may even be true for the RC14 prescription, although both physical and observed fits are consistent with $\beta = -2$. 

Finally, we focus on the physical catalogs and the differences between the prescriptions for $\Delta{a}$. While the fits to the observed planets illustrate how detection biases affect the results and serve as a more direct comparison to previous studies, the fits to the physical planets should be interpreted as the \textit{true} MMEN if one knew of all the planets (within the range probed by the simulations, 0.04 to 0.88 AU).
We find that $\beta$ is steepest for the S14 prescription, comparable between CL13 and 10Hill, and slightly shallower for RC14. All of these MMEN slopes are steeper than the value of $\beta = -1.5$ for the MMSN.
The normalization ($\Sigma_0$) is highest for 10Hill, reflecting that it also tends to be the narrowest definition of $\Delta{a}$ of the four prescriptions; intuitively, a smaller feeding zone implies that a greater solid surface density is necessary to collect the same amount of solid material for forming a planet. In contrast, the approximation used by CL13 ($\Delta{a} = a$) results in the lowest value of $\Sigma_0$. Broadly, these results indicate that the mean MMEN is more massive than the MMSN for the innermost regions, although all of these (averaged) extrasolar disk profiles must cross under the MMSN model at some point due to the steeper $\beta$ (e.g., $\sim 0.3$ AU for CL13 and $\gtrsim 1$ AU for RC14).

\subsection{A diversity of disk profiles: fitting power-laws to individual systems} \label{sec:fit_systems}

\begin{figure}
\centering
\includegraphics[scale=0.45,trim={0.5cm 0.5cm 1cm 0.5cm},clip]{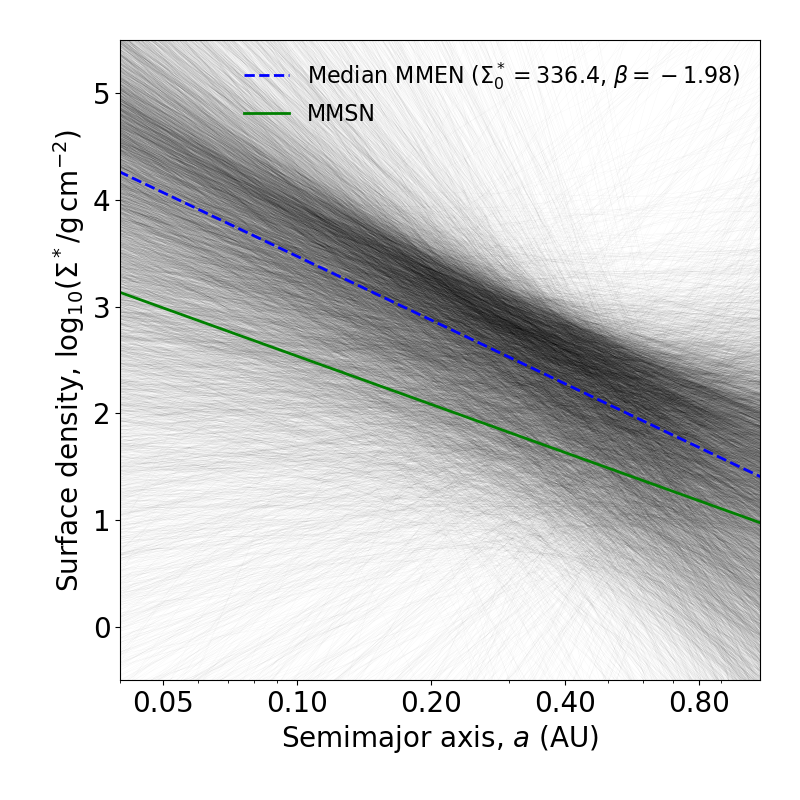}
\includegraphics[scale=0.45,trim={0.5cm 1cm 1cm 1cm},clip]{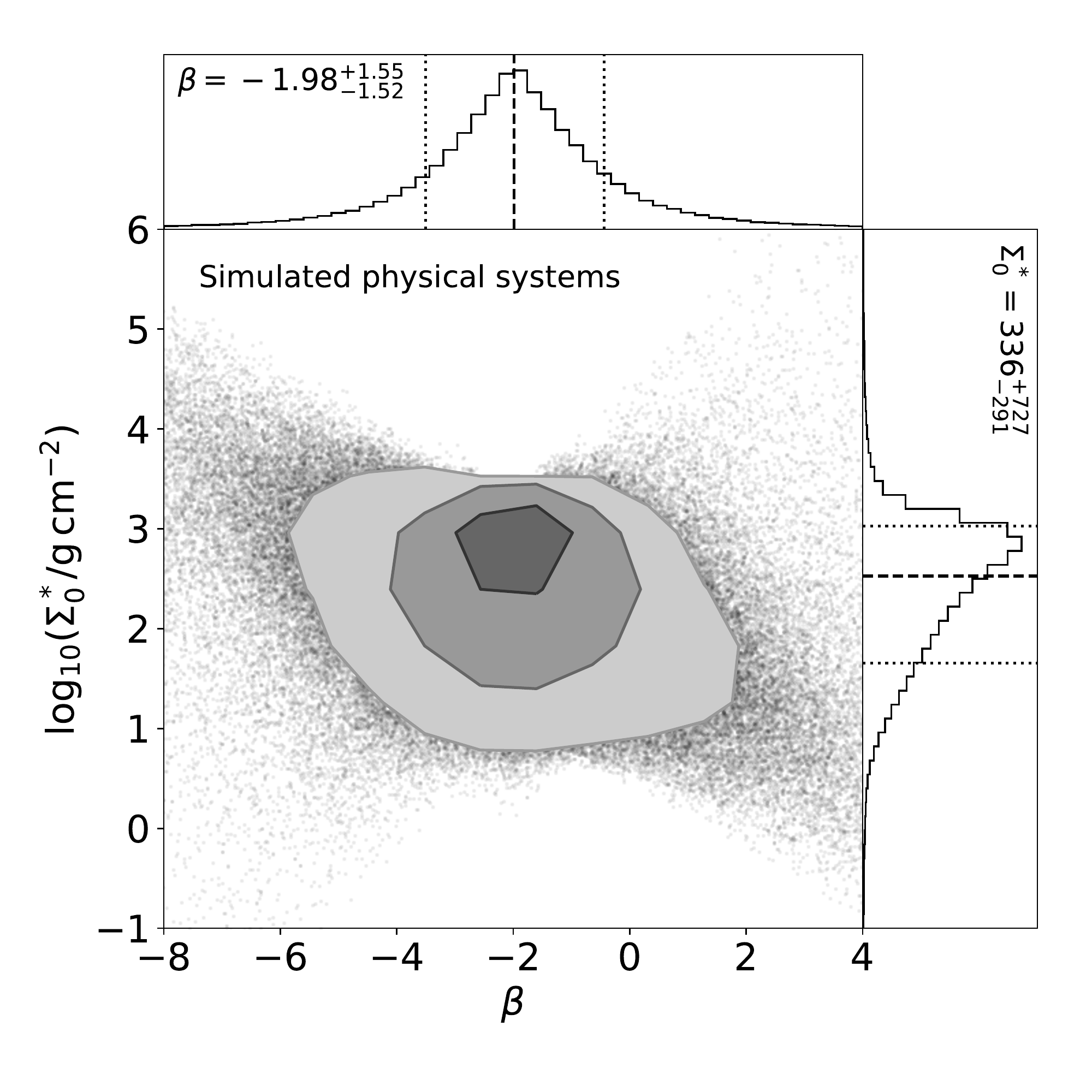}  
\caption{\textbf{Top:} power-law fits to the solid surface densities of planets in each simulated multi-planet system from a physical catalog (a sample of $10^4$ systems are shown). Here, each power-law has been scaled up to be just above the $\Sigma_i$ computed for all planets in the same system; the median scale factor is $\alpha = 2.16$ for systems with at least three planets. The RC14 prescription is used for all planets. \textbf{Bottom:} the corresponding distribution of parameters of the power-law fits to each simulated system from a physical catalog. $\Sigma_0^*$ is the solid surface density at 0.3 AU (again, after scaling by $\alpha$) and $\beta$ is the power-law index on the semi-major axis.}
\label{fig:fit_per_sys_RC14}
\end{figure}

The single power-law model defined above averages over the global population of exoplanets, but it does not describe the properties of any real or single planet-forming disk. Furthermore, it fails to capture the diversity of individual disks. A better approach is given by \citet{2014MNRAS.440L..11R}, who fit the solid surface density profiles of each individual multi-planet system to show that it is inconsistent to assume a universal disk profile. By considering \Kepler{} and RV systems with at least three planets smaller than $5 R_\oplus$ or $30 M_\oplus$, RC14 showed that there is an extremely wide range of slopes ranging from $\beta = -6.3$ to $5.8$.

Here, we adopt a very similar approach to RC14 by also fitting a power-law (equation \ref{eq_mmen}) to each system, starting with the multi-planet (2+) systems in our physical catalog. One issue with this approach is that the resulting power-law for a given system does not guarantee that there is enough solid disk mass to form \textit{every} planet in the system, since by design the power-law fit will be above some points and below the other points in the $\Sigma$ vs. $a$ space (the latter of which have under-predicted local solid surface densities).\footnote{This is not a problem for the two-planet systems, since a power-law can always be fit exactly through both points.} To address this issue, we then ``scale up'' each power-law such that all planets in a given system are at or below the curve, by multiplying $\Sigma_0$ by a scale factor, $\alpha = {\rm max}\{\Sigma_i/\Sigma(a_i)\}$, where $\Sigma_i$ is the solid surface density for the $i^{\rm th}$ planet in the system and $\Sigma(a_i)$ is the solid surface density of the power-law fit evaluated at that planet's semi-major axis.
Hereafter, we will use $\Sigma_0^* \equiv \alpha\Sigma_0$ to denote the scaled-up solid surface density normalization.
This ensures that the resulting power-law profile contains enough mass to form each planet in the system while being self-consistent with the assumption that each planet accreted material from within its feeding zone -- a true ``minimum mass'' extrasolar nebula.

The resulting distribution of power-law fits to each multi-planet system in a physical catalog is shown in Figure \ref{fig:fit_per_sys_RC14} (using the RC14 prescription). In the top panel, we plot a sample of power-laws in $\Sigma$ vs. $a$. The bottom panel shows the corresponding distribution of $\Sigma_0^*$ and $\beta$. While the median MMEN slope ($\beta = -1.98$) is the same as the fit to the full ensemble (Figure \ref{fig:mmen_deltaa}), there is a broad and symmetric distribution with the 16th-84th percentile ranging from $\beta = -3.5$ to $-0.43$. Similarly, there is a wide distribution of $\Sigma_0^*$ (16th-84th percentile ranging from 45 to 1060 g/cm$^2$); the median $\Sigma_0^*$ is also comparable to the fit over all planets considering a scale factor was applied to each individual system (median $\alpha = 2.16$ for systems with 3+ planets) but not to the full ensemble. 
We remind the reader that our values of $\Sigma_0^*$ are normalized at 0.3 AU; projecting the power-laws to 1 AU gives $\Sigma(1\rm AU) = 31.5_{-29.2}^{+142.3}$ g/cm$^2$.
While the extreme ends of the distributions are dominated by fits to systems with just two planets, restricting to 3+ systems still produces a substantially broad distribution: $\beta = -1.96_{-1.17}^{+1.18}$ and $\Sigma_0^* = 427_{-343}^{+642}$ g/cm$^2$.
We find comparable results (a broad diversity of $\beta$ and $\Sigma_0^*$) for the other prescriptions of $\Delta{a}$, as listed in Table 1 under ``Fit each system''.
Remarkably, for all except the S14 prescription, the median slope is highly reminiscent of the predictions from the peas-in-a-pod and pair-wise energy-optimized configurations of planetary systems \citep{2019MNRAS.488.1446A, 2022arXiv220310076W}.

\begin{figure*}
\centering
\includegraphics[scale=0.51,trim={0.8cm 0cm 0cm 0cm},clip]{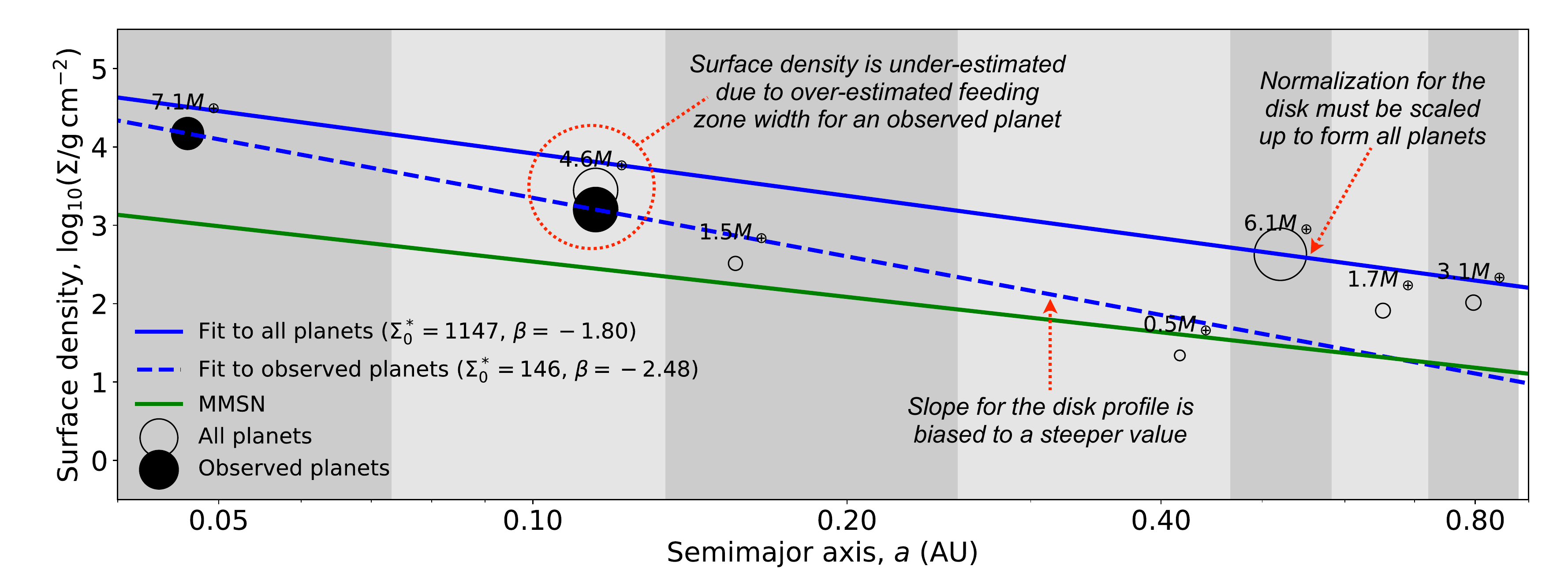}
\includegraphics[scale=0.46,trim={1cm 0cm 0cm 0cm},clip]{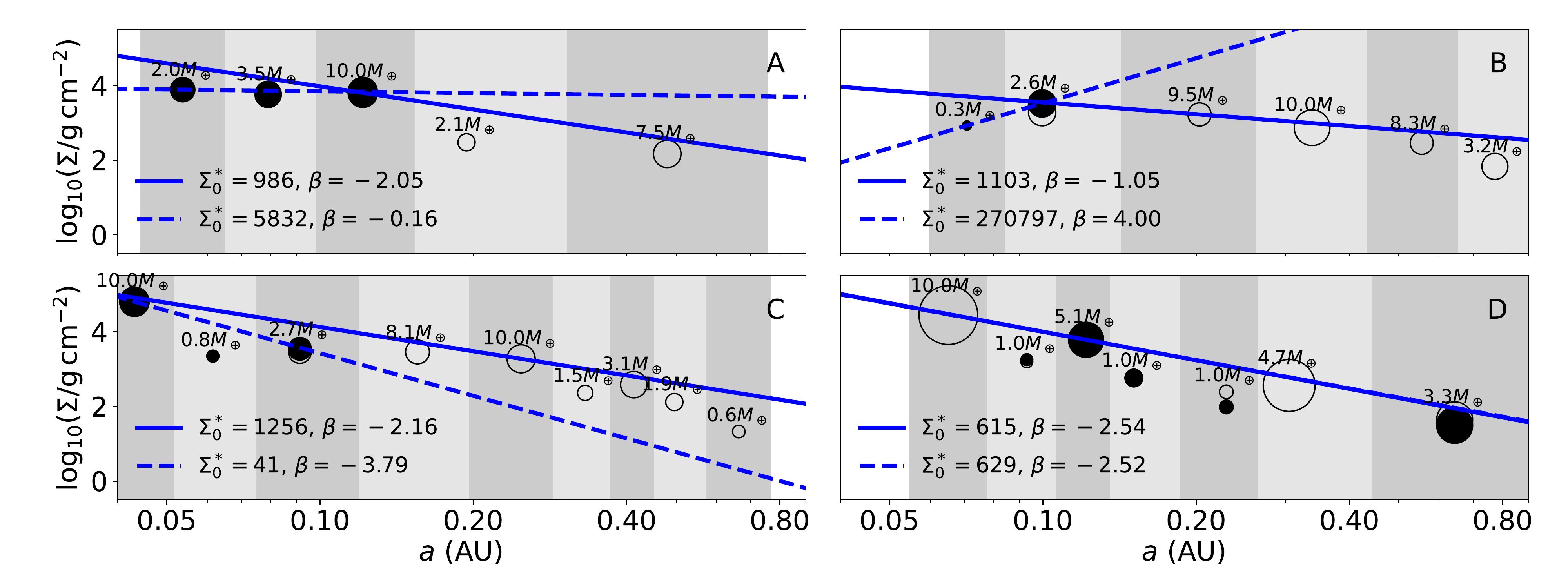} 
\caption{Examples of how missing planets can affect the fitting of MMEN to individual systems. Each panel shows a comparison of a power-law fit to the observed planets only (dashed blue line) versus to all the planets (solid blue line) in a given simulated system. The alternating gray shaded regions denote the feeding zone widths of each individual physical planet, using the RC14 prescription.
\textbf{Top:} an illustration of how undetected planets can alter the inferred disk surface density as demonstrated on a simulated system. Solid and hollow circles that are vertically offset denote the same (observed) planet, which has a biased surface density due to a neighboring missed planet that is causing either an under- or over-estimated feeding zone width. In this case, both $\beta$ and $\Sigma_0^*$ are biased due to the missed outer planets.
\textbf{Bottom:} four simulated systems showing different cases for how the fit to the observed planets can diverge from the fit to the true system. In panel A, $\beta$ is biased to near zero due to two missing outer planets, producing an apparently flat disk profile. In panel B, only the two innermost planets are observed, leading to an extremely positive value of $\beta$ (and $\Sigma_0^*$ which increases arbitrarily towards longer separations) that appears unphysical. In panel C, a large number of undetected planets can also lead to an observed disk profile that is steeper than reality. Finally, panel D shows an example where the inferred disk profile is unaffected despite a missing innermost planet and a middle planet, both of which are among the most massive in the system.
}
\label{fig:fit_examples}
\end{figure*}

\section{The effect of missing planets} \label{sec:missing_planets}

\begin{figure*}
\centering
\includegraphics[scale=0.45,trim={0.5cm 1cm 0.5cm 1cm},clip]{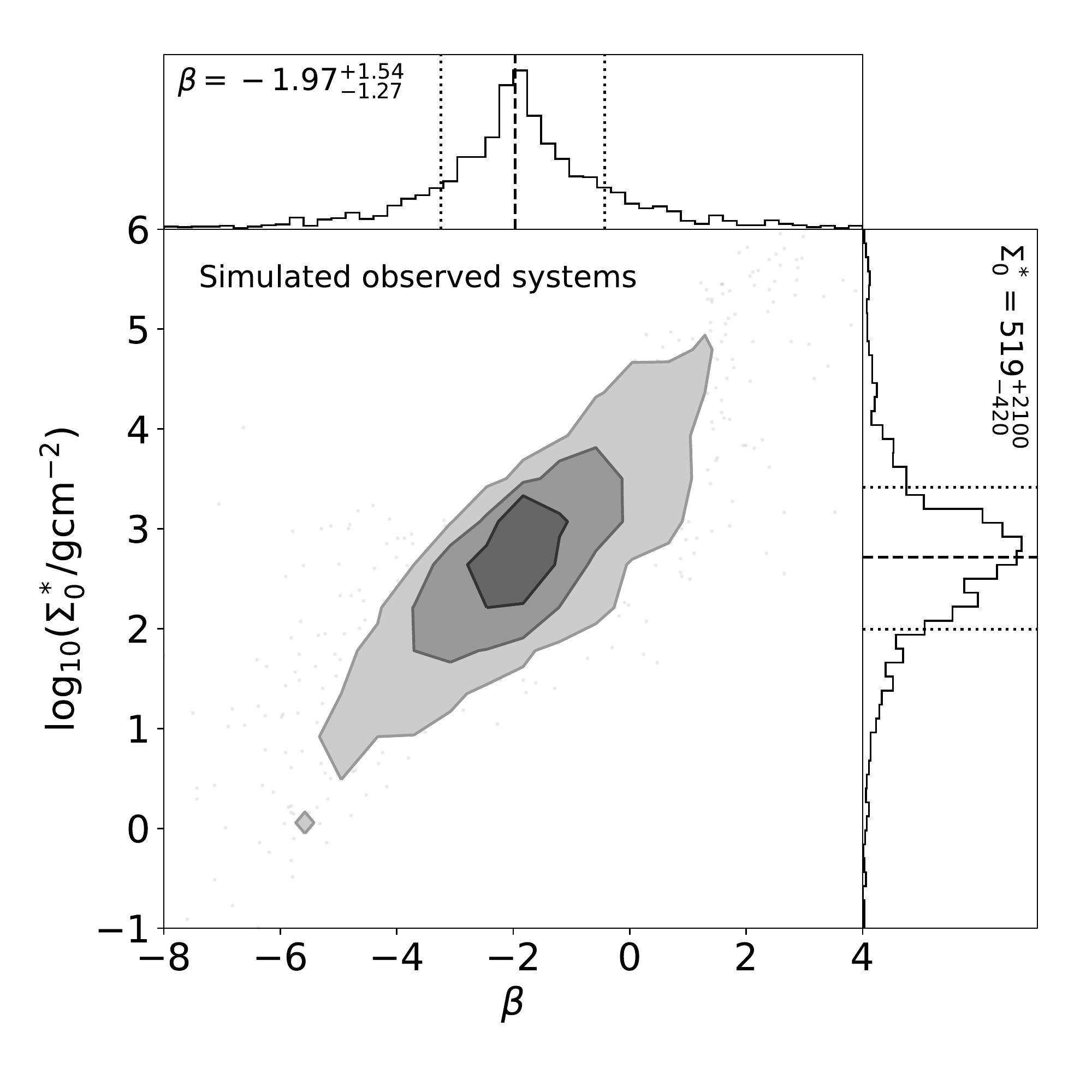}
\includegraphics[scale=0.45,trim={0.5cm 1cm 0.5cm 1cm},clip]{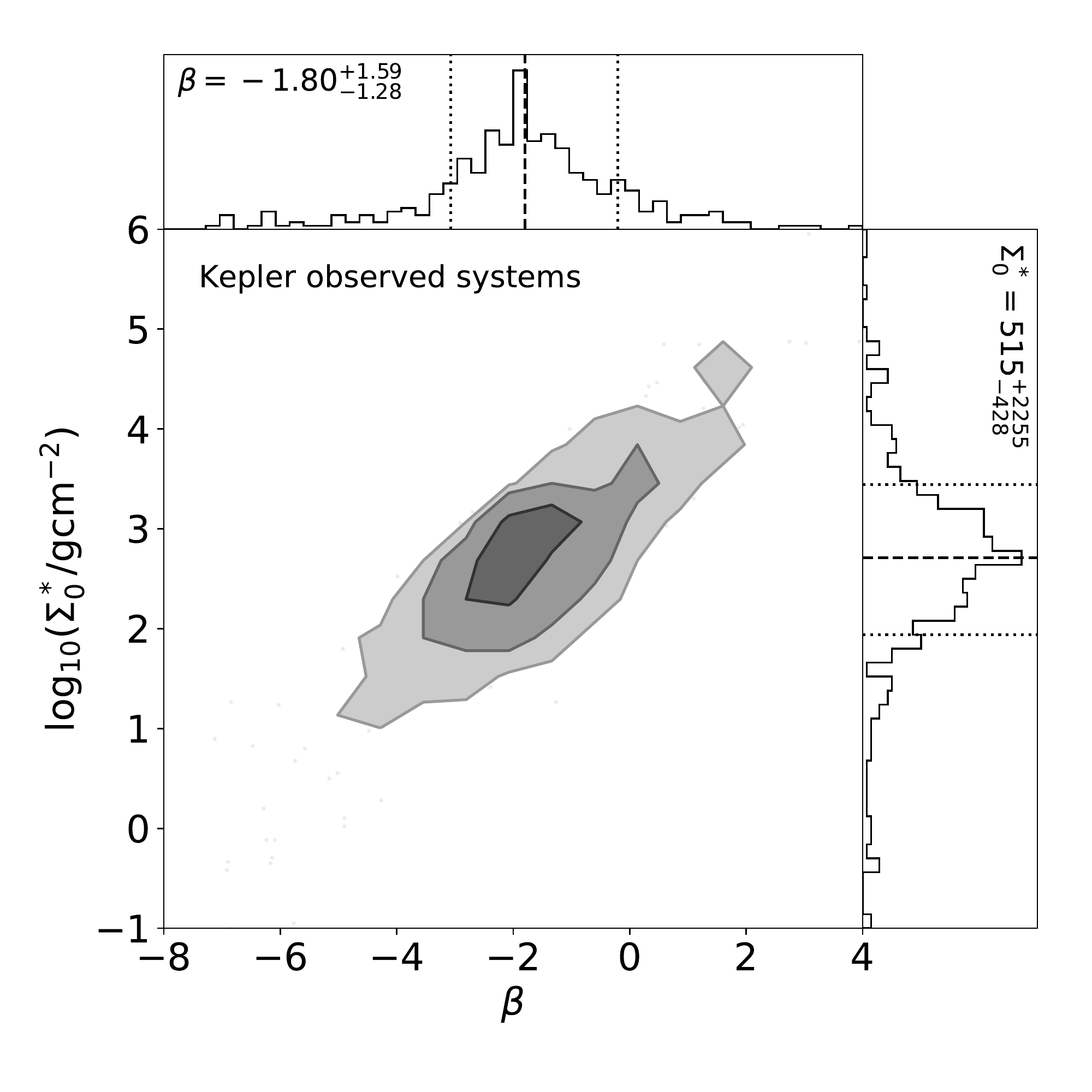} 
\caption{Same as the bottom panel of Figure \ref{fig:fit_per_sys_RC14}, but for a simulated observed catalog (\textbf{left panel}) and the \Kepler{} observed catalog (\textbf{right panel}). In other words, the left panel shows the parameters of the power-law fits to each simulated multi-planet system in an observed catalog (generated from a physical catalog after a \Kepler{}-like mission), while the right panel shows the power-law parameters for \Kepler{} multi-transiting systems. As before, $\Sigma_0^*$ is the solid surface density at 0.3 AU and $\beta$ is the power-law slope on the semi-major axis; the prescription from RC14 is used here for all planets.}
\label{fig:fit_per_sys_obs_RC14}
\end{figure*}

\begin{figure}
\centering
\includegraphics[scale=0.44,trim={0.5cm 0cm 0.5cm 0.2cm},clip]{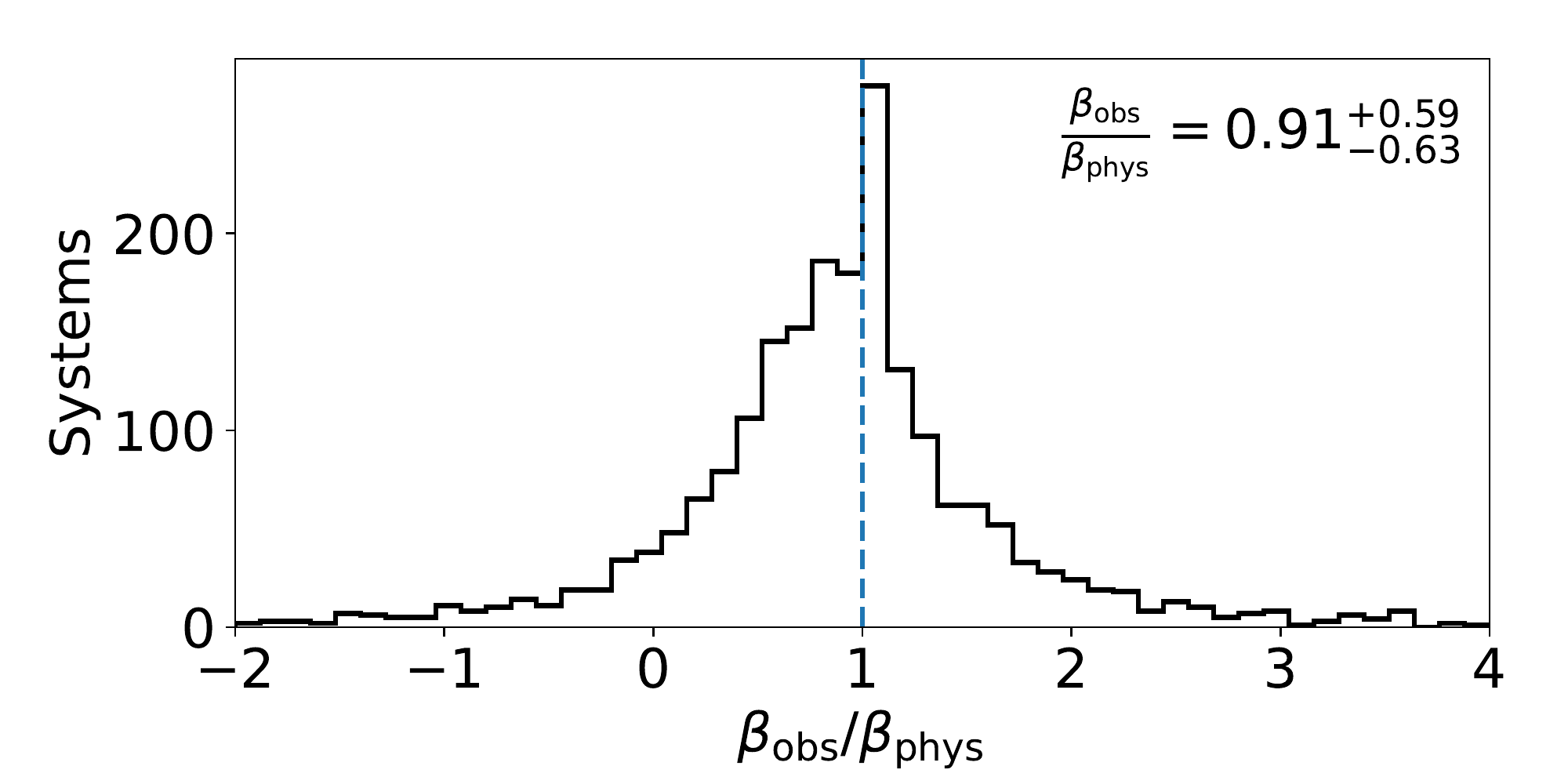}
\includegraphics[scale=0.44,trim={0.5cm 0cm 0.5cm 0.5cm},clip]{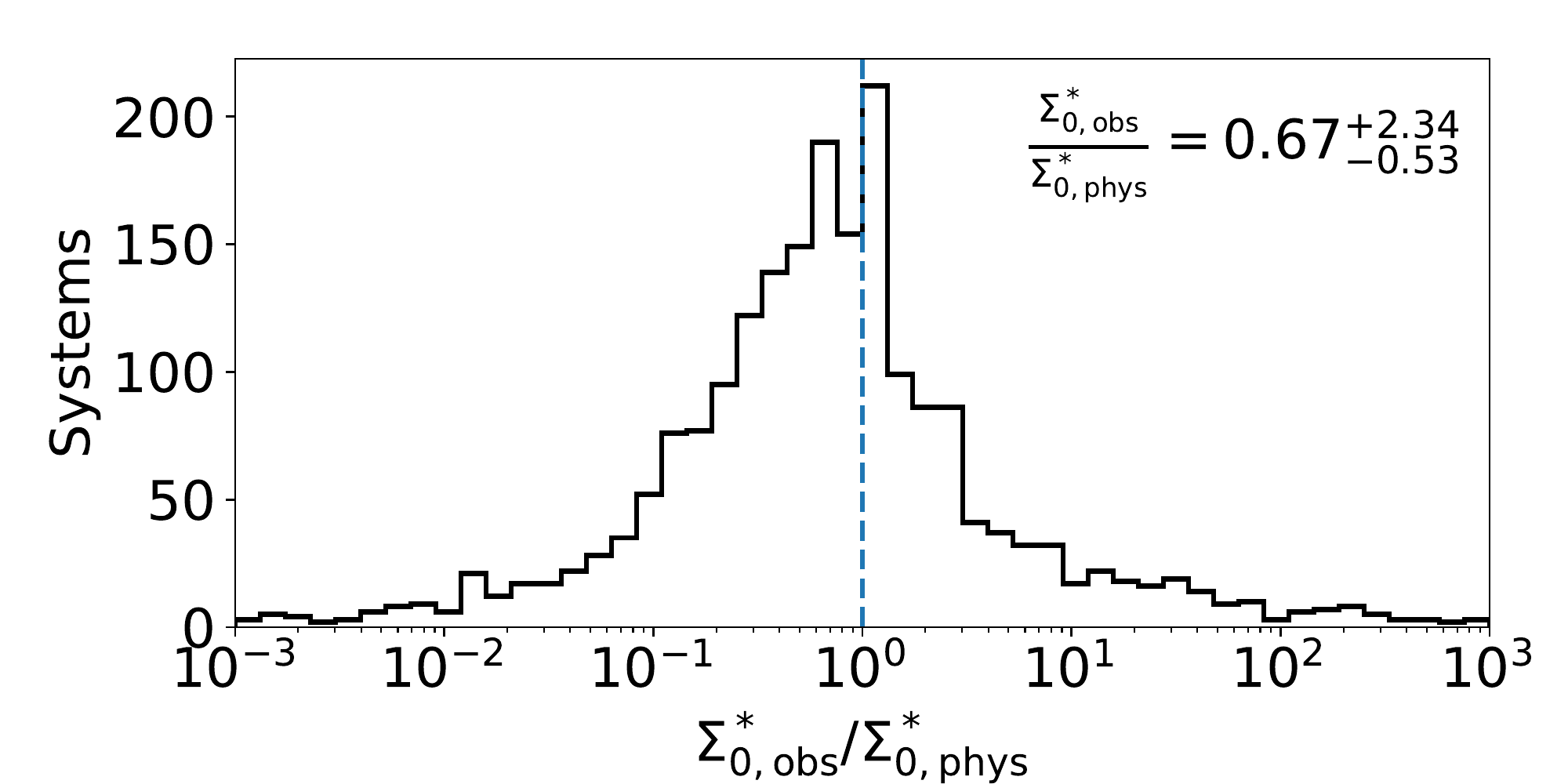} 
\caption{The extent to which the inferred MMEN slope and normalization can be biased due to undetected planets. 
\textbf{Top:} Distribution of slope ratios, $\beta_{\rm obs}/\beta_{\rm phys}$, for the observed systems compared to their true underlying systems. \textbf{Bottom:} Same as the top, but for the normalization ratios, $\Sigma_{0,\rm obs}^*/\Sigma_{0,\rm phys}^*$. In both panels, the vertical dashed line denotes a ratio of one, i.e. unchanged $\beta$ or $\Sigma_0^*$. The listed uncertainties denote the 16th-84th percentiles of each distribution.}
\label{fig:fit_per_sys_obs_vs_phys_RC14}
\end{figure}

In \S\ref{sec:fit_all}, we showed how fitting a power-law to \textit{all} planets in a physical catalog leads to some notable differences in the inferred MMEN compared to fitting to \textit{all} planets in an observed catalog.
While such a comparison illustrates how the overall distribution of planet radii (and thus masses) and semi-major axes is affected by transit detection biases, it does not demonstrate how the biases of a \Kepler{}-like survey affect individual systems, in the form of missing planets due to their non-transiting geometries or undetectably small sizes.
On the other hand, the approach of constructing MMEN from individual systems as described in \S\ref{sec:fit_systems} enables us to directly study this effect, by comparing the power-law fits of the observed systems to those of the physical systems. We note that for the RC14 prescription, the effect of missing planets in a given system is two-fold: (1) the power-law fit (its slope and/or normalization) may be significantly altered or biased, and (2) the assumed feeding zone ($\Delta{a}$) of each planet may also be affected.
We illustrate these biases in Figure \ref{fig:fit_examples} (top panel) using a simulated system. In this particular example, the $4.6 M_\oplus$ observed planet has an under-estimated surface density due to an over-estimated feeding zone width caused by the unseen $1.5 M_\oplus$ planet exterior to it, leading to a disk profile that is biased steep. The power-law fit to the underlying system must also be scaled up to ensure that it can form all the planets, including the $6.1 M_\oplus$ planet ($\alpha \simeq 3.8$ in this case). While this example demonstrates a case in which $\beta$ can appear steeper than reality, a wide range of other cases are also possible, from $\beta$ that are unchanged, shallower, or even positive (bottom panels of Figure \ref{fig:fit_examples}).

In this section, we aim to quantify how these biases affect the overall observed distribution of MMEN disk profiles.
We showed the wide diversity of underlying disk profiles arising from a physical catalog in the previous section. Here, we repeat the procedure of fitting a power-law to each planetary system in the corresponding observed catalog (as well as the \Kepler{}-observed catalog, to test how well our procedure mimics the real Kepler planetary systems). In Figure \ref{fig:fit_per_sys_obs_RC14}, we plot the distribution of power-law fit parameters for each observed multi-transiting system in a simulated catalog and the \Kepler{} catalog, analogous to the bottom panel of Figure \ref{fig:fit_per_sys_RC14}.
The RC14 prescription is used for the figure; the results of the other prescriptions are provided in the bottom half of Table \ref{tab:params} (under ``Fit each system'').

\subsection{Comparing fits to the observed versus physical planets in simulated systems}

Several conclusions can be drawn from comparing the distributions of $\Sigma_0^*$ and $\beta$ fitted between the observed systems (left panel of Figure \ref{fig:fit_per_sys_obs_RC14}) and the physical systems (bottom panel of Figure \ref{fig:fit_per_sys_RC14}).
First, it is remarkable that the distributions of power-law slopes are virtually unchanged between the physical and observed systems; the median $\beta$ and width of the distribution remain similar for each prescription.
This is in contrast to the fits to all planets in a catalog, which tends to lead to slightly shallower slopes for observed catalogs versus physical catalogs for all except the RC14 prescription.
However, we find that the distributions of normalizations are significantly skewed to larger values for all prescriptions. 
While $\Sigma_0^*$ can be either under- or over-estimated for any given system, the median $\Sigma_0^*$ is a factor of $\sim 1.5$ (RC14) to $\sim 2.2$ (CL13) higher for the observed systems than the physical systems. In addition, the spread of the $\Sigma_0^*$ distribution is also substantially increased. This is likely due to the fact that smaller planets tend to be missed more often than larger ones, the latter of which tend to also be more massive, thus biasing the disks inferred from only the observed planets towards higher masses. We note that this is partly counter-acted by the increased feeding zone widths due to missing planets in the RC14 prescription, which explains why the median $\Sigma_0^*$ is least biased using this prescription.

Although the overall distribution of $\beta$ is unaffected by detection biases, both $\beta$ and $\Sigma_0^*$ can be significantly over- or under-estimated for any given system due to missing planet(s), as already illustrated in Figure \ref{fig:fit_examples}. In Figure \ref{fig:fit_per_sys_obs_vs_phys_RC14}, we plot histograms of $\beta$ ratios (top panel) and $\Sigma_0^*$ ratios (bottom panel) for the simulated observed systems, where the ratio is the value of the fit to only the observed planets compared to the fit to all planets in their true underlying systems. The median $\beta$ ratio is close to unity, consistent with the previous comparison showing how the median $\beta$ is insensitive to detection biases. More than half of the observed systems have underestimated values of $\Sigma_0^*$.

RC14 argued against a universal MMEN by showing that synthetic populations of planetary systems generated from a single power-law model (i.e., a fixed value of $\beta$) fail to produce the wide diversity of disk profiles of the \Kepler{}-observed systems even after simplistic simulations of transit detections. While their reported distribution of surface density slopes resulted from fits to the observed systems only, they also showed that a flat distribution of $\beta \in [-2.5, 0]$ or a Gaussian distribution centered around $\beta = -1.25$ (with standard deviation $\sigma = 0.8$) for the underlying distribution appears to roughly match the observed distribution.
While we find a somewhat steeper value for the median surface density slope ($\beta \simeq -2$) compared to RC14, we have demonstrated that this is very insensitive to detection biases when accounting for the role of missing planets using the \citetalias{2020AJ....160..276H} model, and the diversity of slopes across systems remains. 
Additionally, our results show that while detection biases do broaden (and shift) the observed distribution of $\Sigma_0^*$, they cannot account for the full extent of the variance across multi-planet systems. Thus, we further strengthen the conclusion of RC14 that there is no universal MMEN, but rather a diversity of MMEN profiles in the underlying population (Figure \ref{fig:fit_per_sys_RC14}).

\subsection{Comparing fits to the simulated versus \Kepler{} systems}

As seen in Figure \ref{fig:fit_per_sys_obs_RC14}, the distributions of MMEN power-laws are very similar between the \SysSim{} and \Kepler{} observed systems.
There is a systematic shift to slightly shallower median values of $\beta$ for the \Kepler{} systems regardless of the prescription used (e.g. $\sim -2$ vs. $-1.8$ using RC14), similar to the results of fitting a power-law to all planets as discussed in \S\ref{sec:fit_all}.
The distributions of $\Sigma_0^*$ are very similar, although the \Kepler{} catalog appears to have relatively more systems towards the high $\Sigma_0^*$ tail as evidenced by the higher 84\% percentile values for some prescriptions.
Nevertheless, the likeness of the power-law profiles between the simulated systems and the \Kepler{} systems illustrates the robustness of the \citetalias{2020AJ....160..276H} model, and enables us to estimate how missing planets in \Kepler{} systems likely affect our inferences on the MMEN. As we have shown here, the overall distribution of disk surface density slopes is not strongly affected by detection biases and likely represents the \textit{true} underlying distribution. The true disk masses are often higher than what we would conclude from the observed planets alone. In addition, the diversity of disk masses is also reduced for the true systems, although there is still a wide range (as summarized in \S\ref{sec:fit_systems}). In the next section, we use the results from the physical systems to infer the primordial distribution of total disk masses within 1 AU.

\section{Implications for minimum disk masses} \label{sec:disk_masses}

A power-law model for the surface density profile can be integrated to give the total mass in solids enclosed within a radius $r$ from a star:
\begin{align}
 M_r &= \int_{0}^{2\pi}  d\theta \int_{r_0}^{r} \Sigma(a)a\, {da} \\
 &= \begin{cases}
  \frac{2\pi\Sigma_0^*}{(2+\beta){a_0}^\beta} \big(r^{2+\beta} - {r_0}^{2+\beta}\big), &\beta \neq -2 \label{eq_int_mmen} \\
  2\pi\Sigma_0^* {a_0}^2 \ln(r/r_0), &\beta = -2,
 \end{cases}
\end{align}
where $r_0$ denotes the inner edge of the disk where the density must truncate. For $\beta \leq -2$, a non-zero value of $r_0$ is also necessary to avoid an infinite amount of mass. We choose $r_0 \simeq 0.04$ AU, corresponding to a 3-day orbital period around a solar mass star (which is also the minimum period of planets in the \citetalias{2020AJ....160..276H} model).

\begin{figure}
\centering
\includegraphics[scale=0.45,trim={0.5cm 0.5cm 0.5cm 0.5cm},clip]{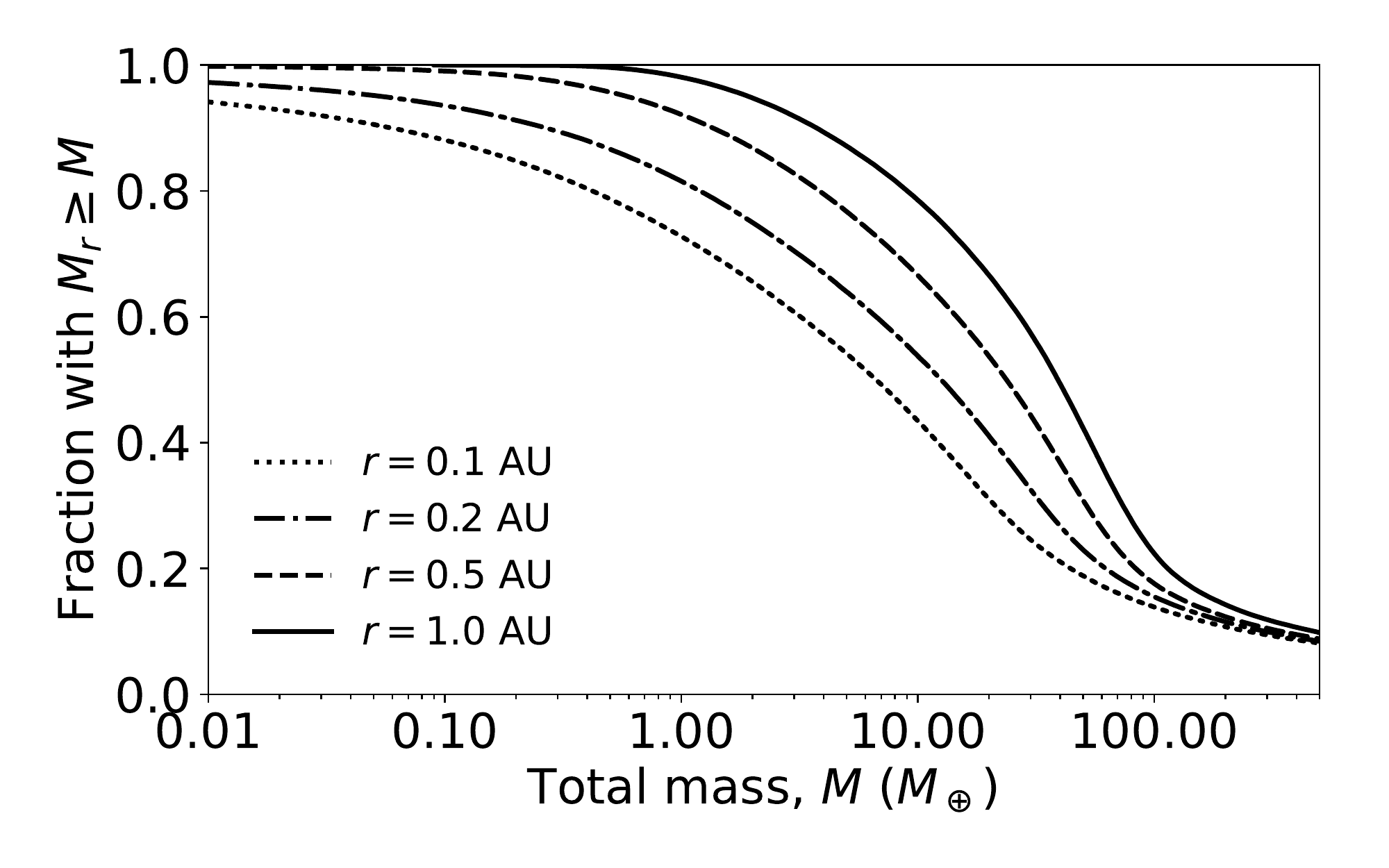} 
\caption{Cumulative distributions of the minimum total mass in solids enclosed ($M_r$) at various separations from a star ($r$, as labeled). The y-axis denotes the fraction of planet-forming disks with \textit{at least} $M$ of solid mass within radius $r$. For example, $\sim 70\%$ of disks have at least $1 M_\oplus$ of solids within 0.1 AU (dotted curve). About half of all disks have more than $40 M_\oplus$ of solids within 1 AU (solid curve).}
\label{fig:total_mass_CDFs}
\end{figure}

We compute $M_r$ for each fitted system in a physical catalog, for several values of $r$, and plot (one minus) the cumulative distributions in Figure \ref{fig:total_mass_CDFs}. The y-axis should be interpreted as the fraction of planet-forming MMEN disks with at least $M_r$ of solid mass enclosed within a given radius $r$.
For example, we find that most such disks have at least an Earth mass of solids within even 0.1 AU (the dotted line). Over 40\% (10\%) of disks have over $10 M_\oplus$ ($100 M_\oplus$) of solids within the same distance. Around a third of disks have over $40 M_\oplus$ within 0.5 AU (the dashed line), and this fraction rises to a half within 1 AU (the solid line). The latter is very similar to the median disk reported by RC14, which also contains $\sim 40 M_\oplus$ within 1 AU (evaluating equation \ref{eq_int_mmen} using their median fit values of $\Sigma(1\, {\rm AU}) = 116$ g/cm$^2$ and $\beta = -1.45$).
The observation that the four curves ($r = 0.1$ to 1 AU) closely approach each other past $M_r \gtrsim 200 M_\oplus$ suggests that for the most massive disks, most of the material would be concentrated in the very inner regions if all the planets formed \textit{in situ} (i.e., these disks have high values of $\Sigma_0^*$ and steep negative values of $\beta$).

\section{Summary and Discussion} \label{sec:discussion}

The MMEN remains an insightful framework for understanding the primordial conditions of protoplanetary disks required to form the planets we see today in their present locations.
Yet despite numerous studies on inferring the MMEN, two key limitations persist: (1) nearly all previous works rely on rudimentary treatments of the transit detection biases,
and (2) most studies (with the exception of RC14) also attempt to fit all known exoplanets with a single power-law model, which does not capture the diversity of planetary systems and the disks from which they formed.
The statistical models for multi-planet system architectures developed via forward modeling of the \Kepler{} mission (e.g., \citealt{2019MNRAS.490.4575H, 2020AJ....160..276H}) provide a detailed and unprecedented way of inferring the MMEN while overcoming both of the above issues.

In this article, we produce various constructions of MMEN using both simulated physical and observed catalogs of exoplanetary systems in addition to the true \Kepler{} catalog, and using a variety of prescriptions for the feeding zone width of each planet (\S\ref{sec:surface_densities}).
First, we follow the widely prevalent approach of fitting a single power-law to each catalog to demonstrate the effect of observational biases on the inferred ``mean'' MMEN and the differences between various prescriptions (\S\ref{sec:fit_all}).
We then fit a power-law to the minimum-mass surface densities of the planets in each planetary system, producing a broad distribution of MMEN (\S\ref{sec:fit_systems}).
By repeating this procedure for individual physical and observed planetary systems, we show that undetected planets in observed systems can significantly alter the inferred disk profile (\S\ref{sec:missing_planets}). Interestingly, while $\beta$ can be strongly affected for any given system, the overall distribution is largely unaffected by missing planets. However, the distribution of $\Sigma_0^*$ is biased and broader for the observed systems compared to the physical systems. Altogether, these results demonstrate that although detection biases do affect the inferred distribution of solid disk profiles, they do not explain all of the variance in the observed profiles. There is no universal MMEN if all planets formed in their present locations.

While our approach is similar to that of RC14, one key difference is that we scale up each power-law fit (i.e., $\Sigma_0^*$ $= \alpha\Sigma_0$ where $\alpha \geq 1$) such that the surface density is no less than that local to any individual planet in the system. This is necessary to ensure that each resulting disk has enough solid mass to form each and every planet while being self consistent with the assumption that each planet accreted solid material from only within its local feeding zone. Our finding that the median $\alpha \simeq 2$ (for systems with 3+ planets) suggests that this consideration alone causes previous studies to typically underestimate the minimum surface densities (and thus minimum disk masses) of planet-forming disks by a factor of two.

RC14 used this wide diversity of MMEN to argue against the \textit{in situ} formation scenario.
Viscous disk models generally predict surface density slopes of $\beta \simeq 0$ to $-2$, depending on the temperature profile (\citealt{1973A&A....24..337S, 1997ApJ...490..368C, 1998A&A...337..625H}; see RC14 for a concise review). Observations of cold dust in disk structures also consistently yield $\beta \simeq -0.4$ to $-1.1$ \citep{2009ApJ...700.1502A, 2010ApJ...723.1241A}. Thus, the wide swath of slopes inferred from applying the MMEN framework to exoplanetary systems, as we have shown, cannot be fully explained by either observations or theory.
Systems with slopes moderately steeper than $\beta \simeq -2$ may be indicative of having witnessed a significant radial drift of solid materials prior to accretion or the migration of planetary embryos or fully-formed planets toward the inner regions of the disks. More sharply falling profiles in the inner regions could be caused by a truncation of the inner planet-forming disk due to the presence of an unknown, distant giant planet. Other extreme slope profiles, including strongly positive values, that exist in the physical systems can potentially be explained by more violent dynamical histories, perhaps involving planetary collisions and/or ejections. Based on the number of simulated physical systems with $\beta < -4$ or $\beta > 0$, we estimate that at least $\sim 23\%$ of planetary systems experienced a history of migration and/or planet-planet interactions that prevent the final planet masses and semi-major axes from conveying information about the initial disk profile. Even among systems with a more typical disk profile, the great diversity of disk masses necessary to enable \textit{in situ} planet formation provides strong evidence that a substantial fraction of these systems experienced a radial drift of solids or substantial orbital migration. In any case, multi-stage formation processes involving the migration, mergers, and scatterings of planetary cores have also been proposed to explain the typical architectures of compact planetary systems (e.g., \citealt{2007ApJ...654.1110T, 2014IAUS..299..360C, 2022arXiv220205342Z}). It appears that a combination of mechanisms beyond the simple \textit{in situ} accretion scenario is necessary to explain all system outcomes.

Additionally, the broad range of inferred solid disk profiles implies that there must have been some very massive or extreme disks, which would likely be unstable. Previous studies have assessed the stability of the gaseous disk by assuming a gas-to-dust ratio (typically $\Sigma_{\rm gas}/\Sigma \sim 200$; e.g. \citealt{2014ApJ...795L..15S}). While outside the scope of this paper, future studies may perform a detailed calculation of the stability of the disks inferred from our \citetalias{2020AJ....160..276H} model using a range of gas-to-dust ratios.

As an alternative to invoking the large scale migration of planets post-formation, it is also possible to consider the radial redistribution of solids (in the form of dust or small pebbles) from which \textit{in situ} planet formation then occurs. Rather than starting from a smooth disk profile, the inward drift of sub-meter sized ``pebbles'' may collect at pressure maxima creating a series of gravitationally unstable rings, from which planets form ``inside-out'' \citep{2014ApJ...780...53C, 2016IAUFM..29A...6T}. The pile-up of solids into narrow annuli from fractions to a few AU which serve as the sites of planet formation can produce steeper radial profiles than the initial disk profile \citep{2016A&A...594A.105D}, and has recently been applied to the inner solar system via the silicate sublimation line at $\sim 1$ AU \citep{2022NatAs...6...72M}.

In theory, one may also evaluate the feasibility of the MMEN model by comparing the distribution of minimum disk masses from the exoplanet population to the distribution of observed protoplanetary disks. Yet, this is challenging for a number of reasons, including the biases inherent to either population and the difficulty of measuring disk masses (see \citealt{2022arXiv220309759D} and \citealt{2022arXiv220309818M} for a review). Observations using the Atacama Large Millimeter/sub-millimeter Array (ALMA) have revealed that Class 0 (young, $<0.5$ Myr) disks typically have $\sim 100 M_\oplus$ of solids while the older, Class I/II disks have less than $\sim 50 M_\oplus$ \citep{2020A&A...640A..19T, 2022arXiv220408731A}. However, these values are sensitive to assumptions for the dust opacity, and extend out to $\sim 100$ AU. On the other hand, ignoring the effects of photons scattering off dust grains can lead to underestimated disk masses \citep{2019ApJ...877L..18Z}. There may also be differences in the stellar samples; ALMA mostly observed disks around K and M dwarfs (e.g., \citealt{2016ApJ...831..125P, 2016ApJ...828...46A, 2018ApJ...869L..41A}), while most exoplanets found by \Kepler{} are around FGK dwarfs. Despite these differences, it has also been suggested that the solid masses of planets and of dust may be similar per star \citep{2021ApJ...920...66M}. It remains that we have a limited understanding of the total solid masses available in the innermost regions of protoplanetary disks, and thus the efficiency of planet formation.

Recent studies have also suggested that the MMEN is dependent on the stellar mass, and to a smaller extent, stellar metallicity \citep{2020AJ....159..247D}. We did not find any correlation between the disk profiles (minimum mass or slope) and stellar mass.
However, it is possible that any underlying correlation is lost due to the scrambling of fitted disk profiles caused by missing planets as we have shown in \S\ref{sec:missing_planets}.
In any case, the inter-system variation in $\Sigma_0^*$ (which can change by a few orders of magnitude) is significantly greater than the range of stellar masses (which at most change by a factor of $\sim 2$ across the FGK range).
A more detailed analysis of how the MMEN varies with host star properties is outside the scope of this paper and should be explored in future work.

The distribution of primordial disk profiles serves as a fundamental input for the initial conditions of planet formation simulations (e.g., \citealt{2013ApJ...775...53H, 2016ApJ...832...34M, 2016ApJ...822...54D, 2020ApJ...891...20M}). It has been shown through simulations of the \textit{in situ} assembly of planetesimals by giant impacts that a diversity of solid disk normalizations can lead to an ensemble of planetary systems resembling the \Kepler{} exoplanetary systems \citep{2020ApJ...891...20M}. However, these studies have typically fixed the surface density slope (e.g., $\beta = -1.5$ or $-2.5$; \citealt{2013ApJ...775...53H, 2016ApJ...832...34M, 2020ApJ...891...20M}) due to the large number of tunable parameters. While it is unlikely that all exoplanetary systems formed exclusively in their present locations, the \textit{in situ} model may still explain a wide variety of planetary system outcomes given an adequately flexible array of initial conditions. The MMEN framework can continue to provide constraints on these formation conditions.

The code for fitting MMEN to individual systems, as well as for reproducing the figures and results of this manuscript, are available via the ``SysSimPyMMEN'' package \citep{matthias_yang_he_2022_7117309}. The simulated catalogs used in this study and code for reproducing the \citetalias{2020AJ....160..276H} model are downloadable from the ``SysSimExClusters'' package \citep{matthias_yang_he_2022_5963884}.

\section*{Acknowledgements}

We thank Darin Ragozzine, Danley Hsu, Robert Morehead, and Keir Ashby for contributions to the broader \SysSim{} project.
We are grateful to Sarah Millholland, Lauren Weiss, and Chao-Chin Yang for helpful discussions.
We also thank the anonymous referee for their constructive review and comments.
M.Y.H. acknowledges the support of the Natural Sciences and Engineering Research Council of Canada (NSERC), funding reference number PGSD3 - 516712 - 2018.
%
%
M.Y.H. and E.B.F. acknowledge support from the Penn State Eberly College of Science and Department of Astronomy \& Astrophysics, the Center for Exoplanets and Habitable Worlds, and the Center for Astrostatistics.  
%
%
%
This research has made use of the NASA Exoplanet Archive, which is operated by the California Institute of Technology, under contract with the National Aeronautics and Space Administration under the Exoplanet Exploration Program.
%
%

\software{NumPy \citep{2020Natur.585..357H},
          Matplotlib \citep{2007CSE.....9...90H},
          ExoplanetsSysSim \citep{eric_ford_2022_5915004},
          SysSimData \citep{eric_ford_2019_3255313},
          SysSimExClusters \citep{matthias_yang_he_2022_5963884},
          SysSimPyPlots \citep{matthias_yang_he_2022_7098044},
          SysSimPyMMEN \citep{matthias_yang_he_2022_7117309}
          }


\bibliographystyle{aasjournal}
\bibliography{main}





\end{document}